# Multifunctional Fiber-based Optoacoustic Emitter for Non-genetic Bidirectional Neural Communication

## Author Information

Nan Zheng[1], Ying Jiang[2], Shan Jiang[3], Jongwoon Kim[3], Yueming Li[4], Ji-Xin Cheng[2,6], Xiaoting Jia[3] * and Chen Yang[5,6] *

## Affiliations

[1] Division of Materials Science and Engineering, Boston University, Boston, MA, USA

[2] Department of Biomedical Engineering, Boston University, Boston, MA, USA

[3] Bradley Department of Electrical and Computer Engineering, Virginia Tech, Blacksburg, VA, USA

[4] Department of Mechanical Engineering, Boston University, Boston, MA, USA

[5] Department of Chemistry, Boston University, Boston, MA, USA

[6] Department of Electrical and Computer Engineering, Boston University, Boston, MA, USA

## Contributions

C.Y. conceived the project. N.Z. and S.J. performed fabrication and characterization of materials. N.Z. and Y.J. performed the stimulation and recording experiments *in vitro* and *in vivo*. N.Z. and Y.L. performed the *in vivo* biocompatibility evaluations. X.J. provided guidance on the multifunctional fiber system. J.X.C. provided guidance on the design of fiber optoacoustic emitter. J.K. provided guidance on optimization of recording and data analysis. The manuscript was written through contributions of all authors. All authors have given approval to the final version of the manuscript.

## Corresponding author

Correspondence to: Chen Yang (cheyang@bu.edu) and Xiaoting Jia (xjia@vt.edu)

**Abstract**

A bidirectional brain interface with both "write" and "read" functions can be an important tool for fundamental studies and potential clinical treatments for neurological diseases. Here we report a miniaturized multifunctional fiber based optoacoustic emitter (mFOE) that first integrates simultaneous non-genetic optoacoustic stimulation for "write" and electrophysiology recording of neural circuits for "read". The non-genetic feature addresses the challenges of the viral transfection required by optogenetics in primates and human. The orthogonality between optoacoustic waves and electrical field provides a solution to avoid the interference between electrical stimulation and recording. We first validated the non-genetic stimulation function of the mFOE in rat cultured neurons using calcium imaging. In vivo application of mFOE for successful simultaneous optoacoustic stimulation and electrical recording of brain activities was confirmed in mouse hippocampus in both acute and chronical applications up to 1 month. Minimal brain tissue damage has been confirmed after these applications. The capability of non-genetic neural stimulation and recording enabled by mFOE opens up new possibilities for the investigation of neural circuits and brings new insights into the study of ultrasound neurostimulation.

**Introduction**

Bidirectional communication with dynamic local circuits inside the brain of individual behaving animals or humans has been an invaluable approach for fundamental studies of neural circuits and for effective clinical treatment of neurological diseases, like epilepsy, Parkinson's disease, and depression[1, 2]. Additionally, bidirectional neural interface paves the way for the closed-loop control, as it could enable more sophisticated, real-time control over neural dynamics[3], behaviors[4] and achieve effective therapeutic effect in neurological disease[5, 6]. To achieve real time assessment of the stimulated outcome, neural interfaces with ability to simultaneously manipulate and directly monitor the neural activities are preferred. Among the technologies developed in past decades, electrical stimulation and electrophysiology recording have been widely used and forms the basis of current implantable devices, which has been applied to clinical applications[7]. For example, to restore both the motor and sensory modalities, electric stimulation of the cortical surface is often associated with electrophysiology recording[8, 9], like electrocorticography (ECoG). Also, the bidirectional electrical stimulation has demonstrated promising treatment effect in neurological diseases, such as epilepsy. The responsive focal cortical stimulation (RNS), leveraging ECoG recording as the trigger to provide stimulation, showed a statistically significantly greater reduction in seizure frequency and the benefits increased over time in a two-year study[10, 11]. However, electrical stimulation has a limited spatial resolution due to current spread. It also interferes with the electrical signals used for recording, leading to "contamination" in electrophysiology recording[2, 12]. Although researchers are improving its performance through technologies such as current steering[13], novel electrode design[14], and artifacts cancellation[15], considering the intrinsic physical properties of brain tissue[16], the current spread, root cause of above-mentioned issues is hard to be fully eliminated. Therefore, electrical stimulation for the bidirectional communication of brain may not be the ideal candidate.

Being orthogonal with electrical recording, optical stimuli not only avoids the interference but also enables a high spatial resolution. To take this advantage, early efforts developed so-called optoelectrodes by simply assembling the optical fibers for optogenetics stimulation with the electrodes, such as Utah arrays[17-19], Michigan probes[20, 21] and microwires[22]. Semiconductor fabrication techniques and multiple material processing methods have recently been applied to improve the integration of those bidirectional devices. New processing techniques not only make the device more compact but also strengthen its functionality and biocompatibility. For example, monolithically integrated micro-light-emitting-diodes (µLEDs) were used to reduce the complexity of light-guide structures and significantly boosted the number of stimulation sites and stimulation resolution [23, 24]. Alternatively, a high-throughput thermal drawing method has been used to integrate the function components, for example, electrodes, microfluidic channels, and optical waveguides, to the flexible multifunctional polymer fiber [25, 26]. Through this approach, the flexible fiber probes showed low bending-stiffness and enabled multifunctionalities, including optogenetics, electrical recording and drug delivery [27-29]. Since optogenetics relies on the expression of light-sensitive opsins in neurons through gene modification[26], it is challenging to apply optogenetics to non-human primates and human effectively and safely[30].

Recently, our team showed non-genetic optoacoustic neural stimulation with a high spatial resolution up to single neuron level[31, 32]. In an optoacoustic process, the pulsed light is illuminated on an absorber, causing transient heating and thermal expansion, and generating broadband acoustic pulses at ultrasonic frequencies[33, 34]. As a light mediated neural modulation method, optoacoustic is an ideal candidate to work with electrical recording for bidirectional neural communication. Compared with existing technologies, it exhibited the advantages as a light mediated method, including a high spatial resolution and minimal crosstalk noise with electrical recording. Importantly, the non-genetic optoacoustic neurostimulation alleviates the challenges and safety concern in optogenetics since no viral transfection is required.

Here, we developed a multifunctional fiber-based optoacoustic emitter (mFOE) as a miniaturized bidirectional brain interface performing simultaneously non-genetic neural stimulation and electrical recording of the neural activities. Through a thermal drawing process,[25, 35] fabrication of mFOE integrated an optical waveguide and multiple electrodes within a single fiber with a total diameter of 300 μm, compatible to the typical size of silica fibers used in optogenetic studies. An optoacoustic coating was selectively deposited to the tip of the core optical waveguide in the mFOE through a controlled micro-injection process. Upon nanosecond pulse laser delivered to the photoacoustic coating, the mFOE generates a peak-to-peak pressure greater than 1 MPa, confirmed by the hydrophone measurement, which is sufficient for successful neural stimulation in vitro and in vivo. By calcium imaging, the optoacoustic stimulation function of the mFOE was validated in Oregon green-loaded rat primary neurons. Importantly, we demonstrated the reliable functions of the chronic implanted mFOE for simultaneously stimulating and recording neurons in mouse hippocampus. Chronic recording also demonstrated that the embedded electrodes could provide long-term neural monitoring with a single-unit resolution. The histological evaluation of the brain tissue response confirmed that our flexible mFOE established a stable and biocompatible multifunctional neural interface. mFOE is the first device integrated both optoacoustic stimulation with electrical recording for bidirectional neural communication. With the bidirectional capabilities and excellent biocompatibility, it offers a non-generic tools probing brain circuits, alternative to the optoelectrode devices, with improved feasibility in non-human primates and human. It also opens up potentials for closed-loop neural stimulation and brain machine interface.

## Results

### Design, fabrication and characterization of mFOE

Towards bidirectional neural communication, we have designed the mFOE to utilize the optoacoustic stimulation as "writing" and electrophysiological recording as "reading" of the neural interface (Fig. 1a). Previously, fiber based optoacoustic emitters have been developed as a miniature invasive ultrasound transducer for the biomedical applications, such as intravascular imaging and interventional cardiology[36, 37]. Recently, our work showed that fiber based optoacoustic emitters can also be applied to neural stimulation in vitro and in vivo, with single neuron resolution and dual site capability[32, 38]. In these studies, typically commercial silica fibers were used, together with optoacoustic coating. However, the silica fiber, with Young's modulus of ~70 GPa, is mismatched with mechanical properties of native neural tissue (kilo- to mega pascals)[2] and not easy to integrate with miniaturized electrodes for recording. In this study, we took advantage of the fiber fabrication method developed by Anikeeva and Yoel[25], and utilized the polymer multifunctional fiber design as the base for the mFOE to delivering nanosecond laser to the optoacoustic coating and to record electrical signals. Specifically, a multifunctional fiber with a core optical waveguide and miniaturized electrodes was fabricated using the thermal drawing process (TDP) as previously reported[27] (Fig. 1b). The waveguide is made of polycarbonate core (PC, refractive index $n_{PC}$ = 1.586, diameter = 150 μm) and polyvinylidene difluoride cladding (PVDF, refractive index $n_{PVDF}$ = 1.426, thickness = 50 μm) as the core and the shell, respectively (Fig. 1c). BiSn alloy is used in surrounding electrodes with diameters of 35 μm because of its conductivity and compatibility with TDP (Fig. 1c). This multifunctional fiber showed broadband transmission across the visible range to near infrared region and sub-megaohm impedance when it has been prepared into two centimetres long[27, 39].

To integrate the optoacoustic converter to the multifunctional fiber, the optoacoustic coating, composed of light absorbers and thermal expansion matrix, is needed to be selectively coated on the core waveguide distal end while keeping the surrounding electrodes exposed and conducting. Compared to previously reported FOE fabrication, here we took several innovative steps. First, a pressure-driven pico-litter injector was used to precisely deposit the optoacoustic materials to the core waveguide distal end. The coating area was controlled through varying the injection volume (0.1 – 0.5 nL), which is controlled by the regulated pressure (2-4 psi) over a set period of time (1-2 s, Supplementary Fig. S1) as described in equation (1),

$$V = C \cdot d_{inner}^3 \cdot p \cdot t \qquad (1)$$

where $V$ is the injection volume, $C$ is a constant attributed to the unit conversion factors, effects of liquid viscosity and the taper angle of micropipette, $d_{inner}$ is the inner diameter of the pico-litter injector, $p$ is the pressure, and $t$ is the deposition time. Two 3D translational stages with stereo microscopes were used to precisely control the deposition localization. Second, instead of using carbon nanotubes (CNT), we used carbon black (CB) embedded polydimethylsiloxane (PDMS) as the composite optoacoustic material. CB exhibited similar wideband light absorption[40], assuring the sufficient photoacoustic conversion for neural stimulation. Importantly, due to its relative low viscosity [41, 42], CB/PDMS composite shows much higher injectability compared with CNT/PDMS, therefore more comparable to the pico-liter deposition process. Through these steps, we successfully coated 10-20 μm thick 10% w/w CB/PDMS composite onto the 150 μm diameter core waveguide distal end while electrodes were still exposed as shown in Fig. 1e. Collectively mFOE with the photoacoustic emitter and multiple electrodes has been successfully fabricated.

To characterize the optoacoustic performance of mFOE, a Q-switched 1030 nm pulsed nanosecond laser was applied with pulse energies of 16.6 μJ, 27.3 μJ and 41.8 μJ, respectively. The generated acoustic waves were measured by a 40 μm needle hydrophone placed at about 100 μm away from the fiber tip. Representative pulse acoustic pulse with a width of approximately 0.08 μs was generated by a single laser pulse as shown in Fig. 1f. Higher input laser pulse energy led to larger acoustic pressure. A peak-to-peak pressure of 1.0, 1.6 and 2.3 MPa were measured with the pulse energy of 16.6, 27.3 and 41.8 μJ, respectively. The frequency spectrum shows the broadband characteristic of typical optoacoustic waves[34], and the peak frequencies are around 12.5 MHz (Fig. 1g). Based on previous work, we expected that such pressure and frequency is capable to successfully stimulate neurons in vitro and in vivo. We also calculated the mechanical index (MI), a commonly used matrix, to evaluate the probability of mechanical damage due to ultrasound generated. The MI of acoustic waves generated by 2.3 MPa is 0.198, lower than 1.9, the safety threshold suggested by the Food and Drug Administration (FDA) safety guidelines.

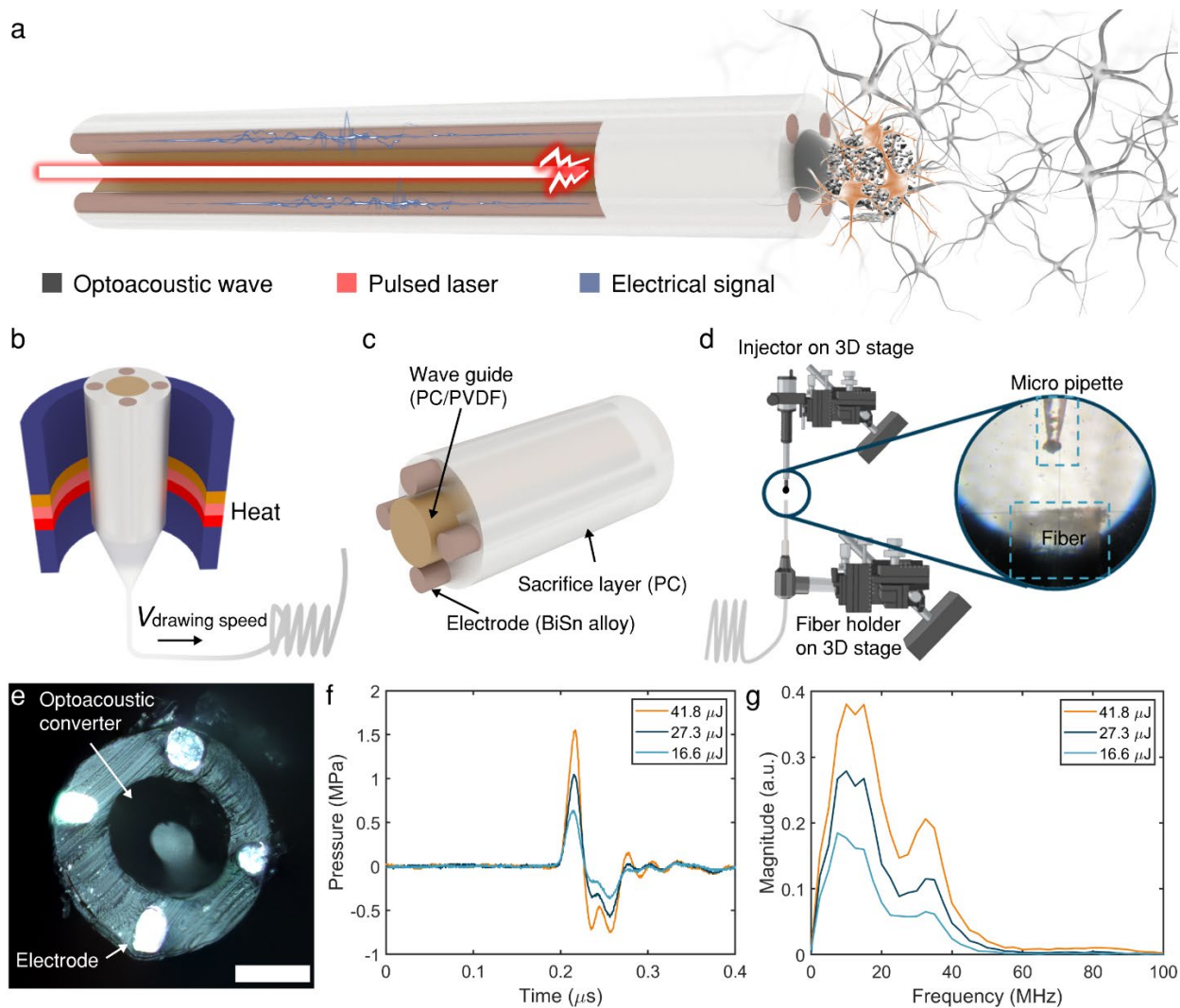

**Figure. 1 Design, fabrication and characterization of mFOE**

**a.** Schematic of mFOE for bidirectional communication with neurons. Input laser pulse (red) is used to generate optoacoustic waves (black) by the converter and the neural activities are recorded by embedded electrodes as the output electrical signal (blue). **b.** Illustration of the thermal drawing process. **c.** Components of the multifunctional fiber, including a PC/PVDF waveguide, BiSn alloy electrodes and PC sacrifice layer. **d.** The selective deposition process for integrating the optoacoustic converter to the core wave guide in the multifunctional fiber. A pressure-driven micro-injector is used to control the volume of CS/PDMS deposited. 3D translation stages and microscope are used to control the deposition location. Zoom-in: The micro pipette was aligned to the center of the fiber under the microscope. **e.** Top view microscope image of the mFOE. Scale bar: 100 µm. **f.** Representative acoustic waveforms under different laser pulse energy recorded by a needle hydrophone. **g.** Frequency spectrum of acoustic waveforms shown in **f**.

## mFOE stimulation of cultured primary neurons

To investigate mFOE can directly trigger the neuronal activity, we examined the response of cultured primary neurons under mFOE stimulation. Because of the presence of calcium channels in neuronal membrane and their activation during the depolarization, calcium imaging has been widely used to monitor the neuronal activities[43, 44]. Here, we cultured and loaded the rat cortical neurons (days in vitro 10-14) with a calcium indicator, Oregon Green™ 488 BAPTA-1 dextran (OGD-1)[45], and performed the calcium imaging with an inverted wide-field fluorescence microscope (Supplementary Fig. S2). To perform the optoacoustic stimulation, mFOE was placed approximately 50 µm above the in-focus target neurons (Fig. 2a) by a micromanipulator under the microscope. 1030 nm 3 ns pulsed laser with a repetition rate of 1.7 kHz was delivered to the mFOE through an optical fiber. The energy of laser pulse was 41.8 µJ, corresponding to a peak-to-peak pressure of 2.3 MPa generated. Lower energy was tested but did not induce calcium transient. The stimulation duration determined by each laser burst was 100 ms, corresponding to 170 pulses (Supplementary Fig. S3). By applying 5 bursts of laser pulses with interval of 1s, we investigated the reproducibility of the stimulation.

Using calcium imaging, we monitored the activities of all neurons in the field of view and divided them into two groups: groups within the converter area (Fig. 2b) and outside the converter area (Fig. 2c). For neurons within the converter area, i.e. the 100 µm from the center of the mFOE, Fig. 2b shows that 8 of 10 neurons showed successful and repeatable calcium transient ($\Delta F/F > 1\%$, the baseline standard deviation) corresponding to each stimulation. Calcium transients are also repeatable for each burst applied over the 1 s period, indicating the evoked neuronal activities and confirming the reliability and biosafety of mFOE stimulation. For neurons outside the converter area, only 2 of 10 neurons responded. This result also suggested the mFOE with the 150 µm center waveguide with photoacoustic coating provided a spatial precision of ~200 µm for stimulation in vitro. This observation is consistent with that fiber based optoacoustic converters generate a confined ultrasound fields with sizes comparable with the radius of converter[31].

Next, to investigate the threshold of mFOE stimulation, we varied the stimulation duration from 5 ms, 50 ms, 100 ms to 200 ms on neurons in different cultures (N = 15) under the same laser pulse energy of 41.8 µJ and the same repetition rate of 1.7 kHz. mFOE stimulation with duration of 5 ms did not evoked any observable fluorescence change (n.s., $p > 0.05$) (Fig. 2g). Only when the duration was 50 ms or longer, the mFOE successfully produced neural activation ($\Delta F/F > 1\%$, $p < 0.01$) as shown in Fig. 2d-f, and Fig. 2h. Longer pulse durations leads to larger peak fluorescence changes, from $2.9 \pm 1.1\%$, $6.0 \pm 2.8\%$ to $7.8 \pm 1.3\%$ corresponding to 50 ms, 100 ms and 200 ms, respectively. For the longest stimulation duration of 200 ms tested, no obvious change on morphology or elevation of baseline fluorescence intensity was detected in neurons after multiple stimulations (Supplementary Fig. 4), indicating the safety of stimulation.

Laser only control experiment was also performed. Laser light with same pulse energy of 41.8 µJ and duration (200 ms, 100 ms and 50 ms) was delivered to OGD-1 loaded neurons through multifunctional fiber without optoacoustic coating. None of neuron culture showed detectable calcium response, distinct from the observed in mFOE stimulated neurons (Supplementary Fig. 5).

To evaluate the photothermal effect of the mFOE stimulation and its potential impact on neurons, we also characterized the thermal profile of the mFOE in PBS during the acoustic generation. Temperature was measured by an ultrafast thermal sensor with a sampling rate of 2000 Hz placed in contact with mFOE optoacoustic coating under the microscope. The laser conditions were consistent with neural stimulation test, i.e., the pulse energy was maintained at 41.8 µJ and the burst duration was varied from 50 ms, 100 ms to 200 ms. The temperature increase on the mFOE surface was found to be $1.23 \pm 0.09$ °C, $1.07 \pm 0.08$ °C, $0.96 \pm 0.08$ °C for 200, 100, 50 ms laser durations, respectively (Supplementary Fig. 6). Such temperature increase is far below the previously reported threshold of thermal-induced neural stimulation ($\Delta T > 5$ °C)[46, 47]. Taken together, we conclude that activation of neurons was due to the mFOE optoacoustic stimulation.

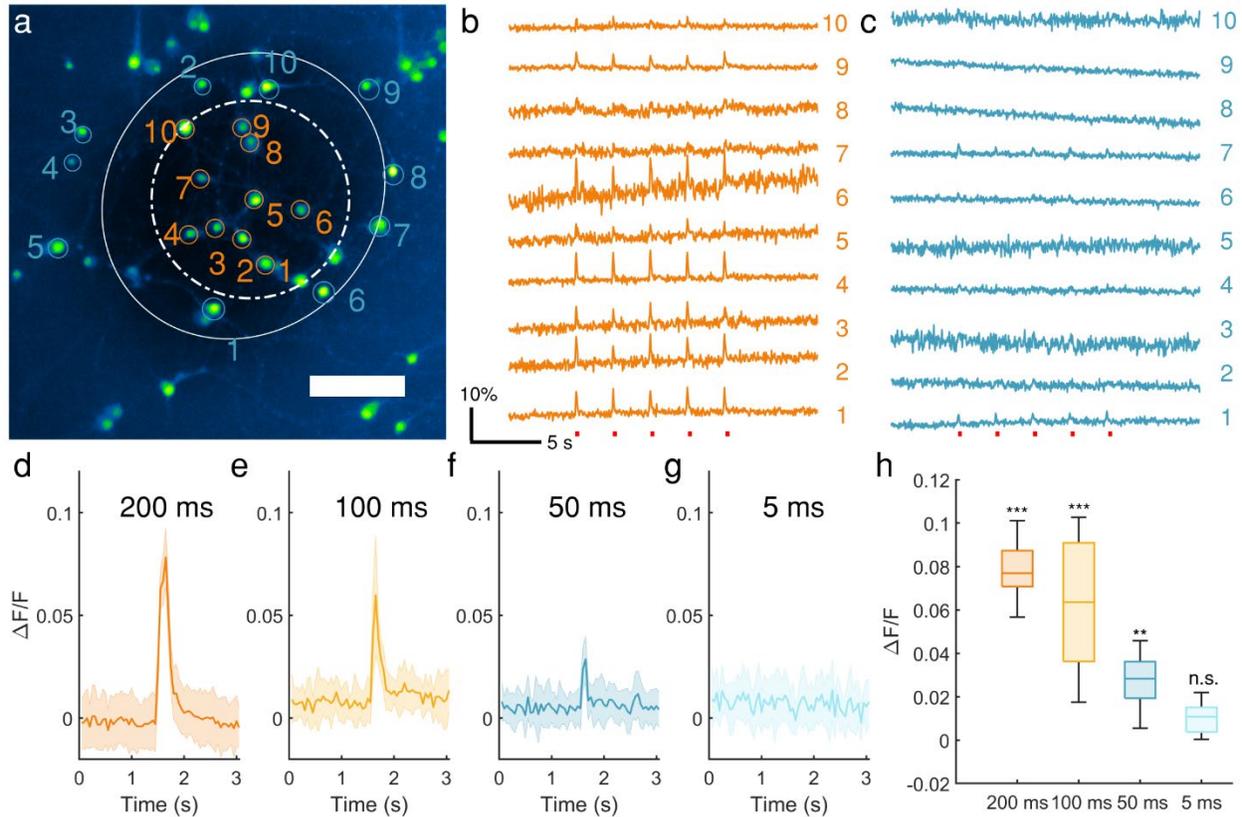

**Figure 2. Calcium transients induced by mFOE in cultured primary neurons.**

**a.** Calcium image of primary cultured neurons loaded with OGD-1. Twenty neurons within (orange) and outside (blue) the optoacoustic converter area are circled and labelled. Scale bar: 100 μm. Solid circle: area outside the converter area; dashed line circle: area within the optoacoustic converter area. **b-c.** Calcium traces of neurons undergone repeated mFOE stimulations with a laser pulse train duration of 100 ms (red dots). Each pulse train was repeated 5 times. Colors and numbers of the traces are corresponding to the neurons labelled in **a**. **d-g.** Average calcium traces of neurons triggered by mFOE stimulation with durations of 200 ms (**d**), 100 ms (**e**), 50 ms (**f**) and 5 ms (**g**), respectively. Shaded area: the standard deviation (SD). N=15 **h.** Average maximum ΔF/F of neurons stimulated by mFOE. N = 15. (n.s.: non-significant, $p > 0.05$; *: $p < 0.05$; **: $p < 0.01$; ***: $p < 0.001$, One-Way ANOVA and Tukey's mean comparison test)

## *In vivo* simultaneous optoacoustic stimulation and electrophysiological recording

Since the animal experiment is a significant part of the study in neuroscience and neurological diseases, we further investigated the performance of mFOE in the wild type *C57BL/6J* mice. In vivo optoacoustic stimulation was performed by delivering pulsed laser to the implanted mFOE, and the optoacoustically stimulated neuronal activities were recorded through electrodes in the mFOE (Fig. 3a). Experimentally, we implanted the mFOE into the hippocampus of mice (N =5). The chronically implanted mFOE allows mice to move freely after surgery (Fig 3b). During stimulation and recording tests, the mFOE was coupled with the laser source and electrophysiological recording headstage through the standard ferrule and pin connector, respectively. The stimulation and recording were conducted in the mice under continuous anesthesia induced and maintained by isoflurane. Based on the threshold of optoacoustic stimulation obtained in *in vitro* studies, 50 ms bursts of laser pulses with a pulse energy of 41.8 µJ were delivered to the mFOE at 1Hz during the 5 second treatment period. The simultaneous electrophysiological recording by mFOE electrodes was bandpass filtered to examine the local field potential (LFP, 0.5-300 Hz). Simultaneous optoacoustic stimulation and electrophysiological recording were performed at multiple time points, including 3 days, 7 days, 2 weeks and 1 month (Fig. 3c-f). Three out of five mice tested showed successful simultaneous stimulation and recording functions for testing periods of 3 days to one month.

The evoked brain activities corresponding to the optoacoustic stimulation were confirmed by monitoring the LFP response. LFP response at two weeks after implantation was detected with latency of $7.19 \pm 2.29$ ms (N = 15, from three mice). The amplitude of LFP response varied at four time points. The largest and smallest responses occurred at 2 weeks and 1 month, respectively. A possible reason for this observation may be the brain tissue injury and healing after the implantation. These results collectively demonstrate the reliability of the optoacoustic stimulation and recording functions of the implanted mFOE in the animals.

To eliminate the possibility that LFP response was induced by electrical noise or laser artifacts, we also conducted two sham control experiments. In the light only control group, we implanted a multifunctional fiber without optoacoustic coating to the mouse hippocampus and delivered the laser light with the same condition. The LFP recorded didn't correlate to the laser pulse train, indicating the spontaneous brain activities were recorded and light only did not invoke the LFP response (Supplementary Fig. 7a). In the dead brain control group, we tested the optoacoustic stimulation through mFOE implanted to the euthanized mouse and did not observe the corresponding LFP response (Supplementary Fig. 7b). These results collectively confirm the signals we detected from mFOE stimulation were not artifacts.

We further evaluated the recording performance of implanted mFOE. To evaluate the ability of mFOE for single unit recording, the electrophysiological signals recorded were bandpass filtered for spike activity (0.5-3 kHz, Fig. 3g). Through a principal-component analysis (PCA) based spike sorting algorithm, two spike clusters can be isolated from an endogenous neural recording (Fig. 3j). The cluster quality was assessed by two common measures[48], $L_{ratio}$ and isolation distance. $L_{ratio}$ is 0.0017 and isolation distance has the value of 99.37. The first averaged spike shape (Fig. 3h) showed a narrower and larger depolarization than that of the second spike shape (Fig. 3i). The different spike waveform and the cluster analysis suggested that the action potentials were recorded from at least two different groups of neurons[49, 50]. Thus, the successfully spike sorted neural activities from CA3 confirmed the ability of mFOE electrodes for the single-unit recording.

To examine the sensitivity of LFP recording, at one month after implantation we altered the anesthesia level via adjusting the induced isoflurane concentration during the recording to see if the characteristic anesthesia dosage-dependent changes can be observed (Fig. 3k). Initially, a low level of anesthesia was maintained at 0.5% v/v isoflurane, and recorded LFP showed that spontaneous brain activities occurred continuously (i in Fig. 3h. and Fig. 3l). Then a higher-level anesthesia (3% v/v isoflurane) was applied for 3 minutes. After increased the isoflurane level, some spontaneous brain activities were suppressed and a hyperexcitable brain state was induced, where the voltage alternation (bursts) and isoelectric quiescence (suppression) appeared quasiperiodically[27, 51] (ii in Fig. 3h and Fig. 3m). With maintaining 3% v/v isoflurane, a deep anesthesia state was induced in the animal. At the same time, both respiration rate and responsiveness to toe pinch decreased due to the higher anesthetic level.

Less voltage alternation occurred and for the most of time the LFP signal was a flat line (suppression, iii in Fig. 3h and Fig. 3n). Compared with initial stage, γ band LFP activity in 30-100 Hz was decreased due to the higher concentration of isoflurane as shown in the power spectrum[52] (Fig. 3n). Later, when the concentration of isoflurane was reduced to 0.5% v/v again, the LFP activity returned to a similar level as measured in the initial stage. Taken together, this isoflurane dosage-dependent characteristic confirmed the accuracy of LFP recording by mFOE.

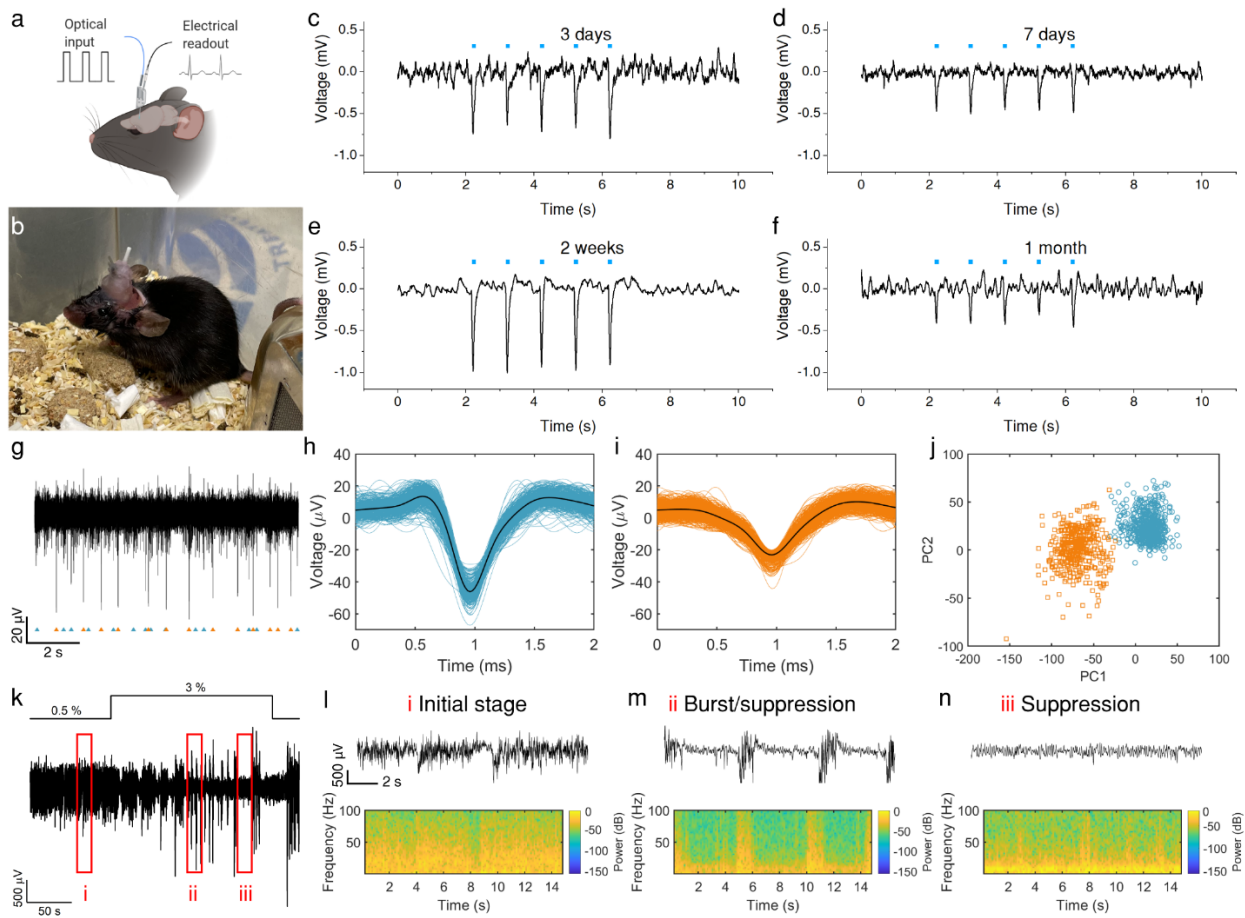

**Figure. 3 Simultaneous optoacoustic stimulation and electrophysiological recording by implanted mFOE in mouse hippocampus.**

**a**. Illustration of the mFOE enabled bidirectional neural communication using laser signal as input and electrical signal as readout. **b.** mFOE was implanted into hippocampus of a wild type C57BL/6J mouse. **c-f.** Simultaneous optoacoustic stimulation and electrophysiological recording performed at 3 days (**c**), 7 days (**d**), two weeks (**e**) and one month (**f**) after implantation. Blue dots the laser pulse trains. For each laser train: 50 ms burst of pulses, pulse energy of 41.8 μJ, laser repetition rate 1.7 kHz. **g.** Part of the filtered spontaneous activity containing two separable units recorded by mFOE electrode at one month after implantation. **h-i.** Spike shapes of two separable units in **g**. **j.** Principal-components analysis (PCA) of the two units. **k.** Local field potential (LFP) recorded by mFOE one month after implantation with an alternating anaesthesia level (0.5-3% v/v isoflurane). **l-n.** different LFP responses induced by varying the concentration of isoflurane: **l** corresponds to the initial stage (0.5% of isoflurane level); **m** corresponds to the burst/suppression transition stage (after increasing the isoflurane level to 3%); **n** corresponds to the suppression stage (the isoflurane level was maintained at 3% and took effect).

# Foreign body response comparison between mFOE and standard optical fiber using immunohistochemistry

Foreign body response is a critical property of implantable neural interface to assure their usage in a safe and chronic way, since the physical insertion into brain tissue commonly initiates a progressive inflammatory tissue response[53]. To evaluate the biocompatibility of mFOE, we compared the foreign body response of mouse brain to mFOE with the similar size standard silica optical fibers (diameter = 300 µm), which is widely used in optogenetic technologies[54, 55]. The immunohistochemistry analysis of surrounding brain tissue was performed from mice (N = 3) implanted with the mFOE and a conventional silica fiber 3 days and 1 month after implantation (Fig. 4a). The damage to surrounding neurons from implant was assessed through evaluating neuronal density using the neuronal nuclei (NeuN) markers (Fig. 4b). Number of neurons was calculated by counting the NeuN-positive cells per field of view (650 × 650 µm). The presence of ionized calcium-binding adaptor molecule 1 (Iba1, Fig. 4c) and glial fibrillary acidic protein (GFAP, Fig. 4d) were used as the markers for activated microglia and astrocytic response, respectively.

    Compared with the silica fiber, mFOE induced significantly less microglial response ($p < 0.01$, Fig. 4c, f) and astrocyte reactivity ($p < 0.001$, Fig. 4d, g), but no significant difference was observed on the neuronal density (Fig. 4b, e) 3 days after implantation. A decrease in foreign body response, specifically, higher neuronal density and lower microglia and astrocytic response (Fig. 4e-g), was observed from 3 days to 1 month after implantation of both mFOE and silica fiber and no significant difference was observed between mFOE and silica fiber 1 month after implantation. Taken together, the immunohistochemistry analysis confirmed that mFOE yielded less foreign body response in the short period, i.e., 3 days, after implantation and showed similar biocompatibilty with silica fiber at longer implantation time, i.e., 1 month.

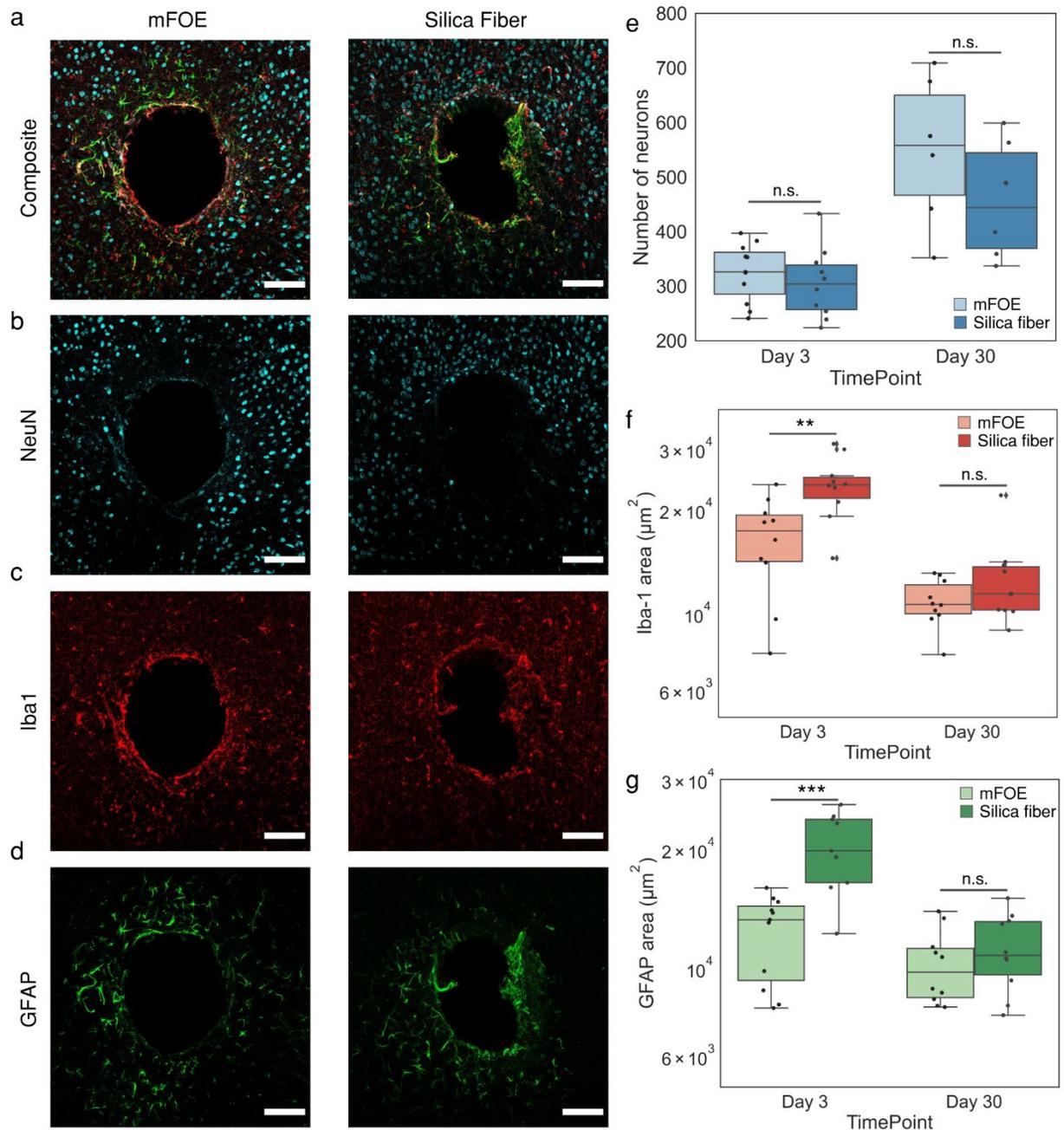

**Figure. 4 Foreign body response comparison of mFOE and silica fiber using immunohistochemistry.**

**a-d.** Immunohistochemistry images of mouse brains implanted with mFOE and silica fiber one month after implantation (N = 3). Scale bar: 100 μm. Brain slices were labelled with the neuron-specific protein (NeuN, cyan), ionized calcium-binding adaptor molecule 1 (Iba1, red) and glial fibrillary acidic protein (GFAP, green). **e.** Number of neurons in the field of view, calculated by counting the NeuN-positive cells for mFOE and silica fiber at 3 days and 1 mon after implantation. **f.** Microglial reactivity, assessed by counting the Iba-1 labelled area, for mFOE and silica fiber at 3 days and 1 mon after implantation. **g.** Astrocyte reactivity, assessed by counting the GFAP labelled area, for mFOE and silica fiber at 3 days and 1 mon after implantation. For each experimental group, two to four brain slices were used from each mouse (N= 3). (n.s.: non-significant, p > 0.05; *: p < 0.05; **: p < 0.01; ***: p < 0.001, One-Way ANOVA and Tukey's mean comparison test)

## Discussion

In this study, we designed and developed a miniaturized fiber-based device, i.e. mFOE, for bidirectional neural communication. mFOE performs the "write" function, i.e. non-genetic optoacoustic stimulation and the "read" function, i.e. simultaneous electrophysiological recording. The broadband acoustic wave with a broadband ultrasound pulse with pulse width about 0.1 μs and a center frequency at 12.5 MHz and a peak pressure of 2.3 MPa with pulse numbers >85 generated by mFOE successfully stimulate neurons with a spatial resolution of approximately 200 μm in primary rat cortical neuron culture. By implanting mFOE into mouse hippocampus, we demonstrated its ability for simultaneous optoacoustic stimulation and electrophysiological recording and superior biocompatibility as a chronic bidirectional neural interface. Reliable stimulation and LFP recording have been achieved up to one month post implantation. Recording quality has been demonstrated by single unit recording.

For the first time, combining this pico-liter deposition and thermal fiber pulling, we successfully integrated an optoacoustic converter to the polymer multifunctional fiber. Different from the conventional dip-coating method[36, 56], the selective deposition through micro-injection allows the easy fabrication of optoacoustic emitter in a volume and position-controlled way. Through the selective deposition, the dimension of optoacoustic emitter is no longer limited by the tip sizes of optical fibers. Our choice of CB/PDMS composite as the optoacoustic material is also essential as it is comparable with this deposition process with a fine volume control at pico liter level. Besides the application in neural interface, such design and fabrication method can also be applied to optical ultrasound probes used in imaging[37, 57], for example, in the tip engineering and the integration to photonics crystal fibers.

We introduced the optoacoustic stimulation as a new strategy for "writing" in the bidirectional neural interface. Compared with previous optoelectrode devices based on optogenetics[24, 25, 27] and photothermal[58, 59], the non-genetic optoacoustic stimulation enabled by mFOE reduces the barrier of transgenic techniques for applications in primate and potentially human, and avoids the thermal toxicity. At the same time, it offers the spatial precision benefit from the confined ultrasound field. It is orthogonal to electrical recording, therefore minimizing crosstalk with electrical recording. As an emerging neuromodulation method, the mechanism of optoacoustic stimulation is still not fully understood but more studies indicated that mechanosensitive ion channels are responsible for the activation of neurons[60, 61].

Bidirectional brain interfaces are important research tools to understand brain circuits, potential treatments for neurological disease and bridges to brain computer interface for broad applications. New features of mFOE compared to the previous fiber based interface, such as non-genetic and non-electrical stimulation are critical to advance these applications. For example, closed-loop neuromodulation has been demonstrated to be superior to the conventional open-loop system, as it can achieve more responsive and real-time control over neural dynamics. In neurological diseases treatment, combining the detection and in situ intervention improves the treatment effectiveness and safety. Because of its bidirectional capabilities, mFOE has the potential to be used as a new brain interface with closed-loop capability. Using epilepsy as an example, by implanting the mFOE into seizure foci, the continuous LFP recording can guide the localized optoacoustic stimulation and intervene can be triggered at the early stage before seizure progresses into a generalized seizure. The unique orthogonal non-electrical optoacoustic stimulation and electrical recording prevents "contamination" of the recording signals, potentially offering a more effective closed-loop strategy.

In comparison of the optoelectrodes fabricated through semiconductor fabrication process, the recording and stimulation sites of the current mFOE design is fixed at the core waveguide and the number of channels is limited because of the nature of multifunctional fiber. Some post processing methods have been proposed to tackle this challenge, like the laser micromachining technique[27]. In addition, it is possible to further engineer the fiber to offer multiple and selective stimulation sites[62]. With the further development of multifunctional fiber strategy, we believe the bandwidth of mFOE would be improved and open more opportunities in the research of neuroscience and neurological diseases.

## Methods

### Multifunctional fiber fabrication and optoacoustic emitter integration

Multifunctional fibers were fabricated from a preform fiber and then drawn into thin fibers through TDP in a customized furnace. For the preform fiber, PVDF film (Mcmaster) and PC film (laminated plastics) were rolled onto a PC rod (Mcmaster) and followed by a consolidation process in vacuum at 200 °C. Next, four rectangular grooves (2 mm × 2 mm) were machined on the solid PC layer and inserted with the BiSn (Indium Corporate) electrodes. Then, another PVDF layer was rolled over the rod to form an insulation layer for the electrodes and followed by an additional PC as the sacrifice layer for the convenience of TDP. The detailed fabrication process was discussed in the previous paper[27].

A composite of 10% carbon black (diameter < 500 nm, Sigma Aldrich) and 90% polydimethylsiloxane (PDMS, Sylgard 184, Dow Corning Corporation, USA) were used as the optoacoustic material. The mixture was sonicated for 1 hour followed by degassing in vacuum for 30 minutes. The mixture was then filled in the glass micropipette (Inner diameter = 30 μm, TIP30TW1, World Precision Instruments, USA) connected to the pico-liter injector (PLI-100A, Warner Instruments, USA). Under the microscope, the glass micropipette was aligned with the core waveguide of multifunctional fiber and the mixture was deposited to the surface of the core waveguide by controlling the injection pressure and time. The deposited fiber was then cured vertically at room temperature for 2 days.

Before use, mFOE was further prepared for the optical coupling and electrodes connection. For the optical coupling, a ceramic ferrule (Thorlabs, USA) was added and affixed to the end of the fiber by the 5-min epoxy (Devcon, ITW Performance Polymers, USA). Then the end surface was polished by optical polishing papers to reduce roughness from 30 μm to 1 μm. For the connection to electrodes embedded in the multifunctional fiber, the electrodes were exposed manually along the side wall of the fiber by using a blade and silver paint (SPI Supplies, USA). Then copper wires were wrapped around the fiber at each exposure locations along the fiber and the silver paint were applied for the fixation and lower resistance. The copper wires connected to fiber electrodes were soldered to the pin connector while a stainless-steel wire was also soldered as the ground wire for later extracellular recording. In addition, the 5-min epoxy (Devcon, ITW Performance Polymers, USA) was applied to the connection interface for strengthening affixation and better electrical insulation.

## Optoacoustic wave characterization

To generate the optoacoustic signal, a compact Q-switched diode-pumped solid-state laser (1030 nm, 3 ns, 100 μJ, repetition rate of 1.7 kHz, RPMC Lasers Inc., USA) was used as the excitation laser source. The laser was first connected to an optical fiber through a 200 μm fiber coupling module and then connected to the mFOE with a SubMiniature version A (SMA) connector. The pulse energy was adjusted through a fiber optic attenuator (varied gap SMA Connector, Thorlabs, Inc., USA). The acoustic signal was measured through a homebuilt system including a needle hydrophone (ID. 40 μm; OD, 300 μm) with a frequency range of 1–30 MHz (NH0040, Precision Acoustics Inc., Dorchester, UK), an amplifier and an oscilloscope. The mFOE tip and hydrophone tip were both immersed in degassed water. The pressure values were calculated based on the calibration factor provided by the hydrophone manufacturer. The frequency data was obtained through a fast Fourier transform (FFT) calculation using the OriginPro 2019.

## Embryonic neuron culture

All experimental procedures complied with all relevant guidelines and ethical regulations for animal testing and research established and approved by Institutional Animal Care and Use Committee (IACUC) of Boston University (PROTO201800534). Primary cortical neurons were isolated from embryonic day 15 (E15) Sprague−Dawley rat embryos of either sex (Charles River Laboratories, MA, USA). Cortices were isolated and digested in TrypLE Express (ThermoFisher Scientific, USA). Then the neurons were plated on poly-D-lysine (50 μgmL−1, ThermoFisher Scientific, USA)-coated glass bottom dish (P35G-1.5-14-C, MatTek Corporation, USA). Neurons were first cultured with a seeding medium composed of 90% Dulbecco's modified Eagle medium (ThermoFisher Scientific, USA) and 10% fetal bovine serum (ThermoFisher Scientific, USA) and 1% GlutaMAX (ThermoFisher Scientific, USA), which was then replaced 24 h later by a growth medium composed of Neurobasal Media (ThermoFisher Scientific, USA) supplemented with 1× B27 (ThermoFisher Scientific, USA), 1× N2 (ThermoFisher Scientific, USA), and 1× GlutaMAX (ThermoFisher Scientific, USA). Half of the medium was replaced with fresh growth medium every 3 or 4 days. Cells cultured in vitro for 10−14 days were used for Oregon Green labelling and PA stimulation experiments.

### *In vitro* neurostimulation and calcium imaging

Oregon Green™ 488 BAPTA-1 dextran (OGD-1) (ThermoFisher Scientific, USA) was dissolved in 20% Pluronic F-127 in dimethyl sulfoxide (DMSO) at a concentration of 1 mM as stock solution. Before imaging, neurons were incubated with 2 μM OGD-1 for 30 min, followed by incubation with normal medium for 30 min. Q-switched 1030 nm nanosecond laser was used to generate light and delivered to mFOE. The pulse energy was adjusted through a fiber optic attenuator (varied gap SMA Connector, Thorlabs, Inc., USA). Notably, 1030 nm is far from the excitation peak of Oregon Green (494 nm) and pass band of emission filter (500-540 nm), therefore assuring no effect from direct excitation of OGD by any light leak from the fiber. A 3D translational stage was used to position the mFOE approaching the target neurons.

Calcium fluorescence imaging was performed on a lab-built wide-field fluorescence microscope based on an Olympus IX71 microscope frame with a 20× air objective (UPLSAPO20X, 0.75NA, Olympus, USA), illuminated by a 470 nm LED (M470L2, Thorlabs, USA), an emission filter (FBH520-40, Thorlabs, USA), an excitation filter (MF469-35, Thorlabs) and a dichroic mirror (DMLP505R, Thorlabs, USA). Image sequences were acquired with a scientific CMOS camera (Zyla 5.5, Andor,

Oxfords Instruments, UK) at 20 frames per second. The fluorescence intensities, data analysis, and exponential curve fitting were analyzed using ImageJ (Fiji) and MATLAB 2022.

## Implantation surgery procedure

All surgery procedures complied with all relevant guidelines and ethical regulations for animal testing and research established and approved by Institutional Animal Care and Use Committee (IACUC) of Boston University (PROTO201800534). Eight to ten weeks old male wildtype C57BL/6-E mice (Charles River Laboratories, US) were received and allowed to acclimate for at least 3 days before enrolling them in experiments. All mice in experiments had access to food and water *ad libitum* and were kept in the BU animal facility maintained for 12-h light/dark cycle. During the implantation surgery, mice were anesthetized by isoflurane (5% for induction, 1-3.5% during the procedure) and positioned on a stereotaxic apparatus (51500D, Stoelting Co., USA). After hair removal, a small incision was made by sterile surgery scalpel at the target region and then a small craniotomy was made by using a dental drill. Assembled mFOE was inserted into mice hippocampus (−2.0 mm AP, 1.5 mm ML, 2 mm DV) using the manipulator with respect to the Mouse Brain Atlas. The ground stainless steel wire was soldered to a miniaturized screw (J.I. Morris) on the skull. Finally, the whole exposed skull area was fully covered by a layer of Metabond (C&B METABOND, Parkell, USA) and dental cement (51458, Stoelting Co., USA). Buprenorphine SR was used to provide long effective analgesia after the surgery.

## In vivo electrophysiology recording and optoacoustic stimulation

Extracellular recording was performed through an electrophysiology system (Molecular Devices, LLC, USA). mFOE electrodes were connected to the amplifier (Multiclamp 700B, Molecular Devices, LLC, USA) through the pin connector and headstages after the animals recovered from surgeries. The amplified analog signal was then converted and recorded by the digitizer (Digidata 1550, Molecular Devices, LLC, USA).

Q-switched 1030 nm nanosecond laser was used to generate light and delivered to mFOE. During the extracellular electrophysiological recording, the preset trigger signal was generated by the digitizer and used to trigger the Q-switch laser for optoacoustic stimulation. The pulse energy was adjusted through a fiber optic attenuator (varied gap SMA Connector, Thorlabs, Inc., USA).

Data analysis was performed with Matlab and OriginPro and custom scripts were used to analyse the local field potential and spike sorting. The raw extracellular recordings were first band filtered for local field potential results (LFP, 0.5 – 300 Hz) and spike results (300 – 5000 Hz). A custom Matlab script was used to create spectrograms to visually support the analysis of the LFPs in both the time domain and the frequency domain. The spike sorting algorithm was implemented through several steps: first, individual spike signals with length of 3 ms were picked up from the full recording through a standard amplitude threshold method; then the dimensionality of each spike signal was reduced via the principal component analysis (PCA) and unsupervised learning algorithms (K-means clustering) was used to separate out the clusters.

## Foreign body response assessment via immunohistochemistry

To compare the tissue response, animals were implanted with a silica optical fiber (diameter = 300 μm, FT300EMT, Thorlabs, Inc, USA) and mFOE for 3 days or 4 weeks. Then at target timepoints, animals were euthanized and transcardially perfused with phosphate-buffered saline (PBS, ThermoFisher Scientific, USA) followed by 4% paraformaldehyde (PFA, ThermoFisher Scientific, USA) in PBS. The fiber probes were carefully extracted before the extraction and then the brains were kept in 4% PFA solution for one day at 4 °C. Brains were sectioned in the horizontal plane at 75 μm on a vibrating blade vibratome. Free-floating brain slices were washed in PBS and blocked for 1 hour at room temperature in a blocking solution consisting of 0.3% Triton X-100 (vol/vol) and 2.5% goat serum (vol/vol) in PBS. After blocking, brain slices were incubated with the primary antibodies in the PBS solution with 2.5% goat serum (vol/vol) for 24 hours at 4 °C. Primary antibodies used included rat anti-GFAP (Abcam Cat. #

ab279291, 1:500), chicken anti-NeuN (Millipore Cat. # ABN91, 1:500), and rabbit anti-Iba1 (Abcam Cat. # ab178846, 1:500). Following primary incubation, slices were washed three times with PBS for 10 min at room temperature. The brain slices were then incubated with secondary antibodies in the PBS solution with 2.5% goat serum (vol/vol) for 2 hours at room temperature. Secondary antibodies used included goat anti-rat Alexa Fluor 488 (Abcam Cat. # ab150157, 1:1000), goat anti-rabbit Alexa Fluor 568 (Abcam Cat. # ab175471, 1:1000) and goat anti-chicken Alexa Fluor 647 (Abcam Cat. # ab150171, 1:1000). Slices were then washed three times with PBS for 10 min at room temperature. Before imaging, slices were stained with DAPI solution (1 μg/ml, Millipore, USA) for 15 minutes at room temperature. All fluorescent images were acquired with a laser scanning confocal microscope (Olympus FV3000) with an air 20× objective with a numerical aperture NA = 0.75 unless otherwise noted. Neuron density was then calculated within the normalized area by counting NeuN labeled cell bodies using the cell counter plugin (ImageJ). Area analysis of Iba1 and GFAP labeled cells was performed by creating binary layers of the fluorescence images using the threshold function and quantified using the measurement tool (ImageJ).

## Statistical information

Data shown are mean ± standard deviation. For the comparison on peak fluorescence change of in vitro optoacoustic stimulation, one-way ANOVA and Tukey's mean comparison test were conducted by using OriginLab. 15 stimulation events were compared for each condition. For the comparison of foreign body response between silica fiber and mFOE, $N > 8$ brain slices from 3 animals were analysed using one-way ANOVA and Tukey's mean comparison test. The p values were determined as n.s.: nonsignificant, $p > 0.05$; *: $p < 0.05$; **: $p < 0.01$; ***: $p < 0.001$. Statistic analysis were conducted using OriginPro.

## Data Availability

The raw data that support the findings of this study are available from the corresponding author upon request.

## Code Availability

The MATLAB scripts for analysis are available from the corresponding author upon request.

## Acknowledgements

This work was supported by National Institute of Health Brain Initiative R01 NS109794 to J-XC and CY. Research reported in this publication was supported by the Boston University Micro and Nano Imaging Facility and the Office of the Director, National Institutes of Health of the National Institutes of Health under award Number S10OD024993

# Supplementary Information

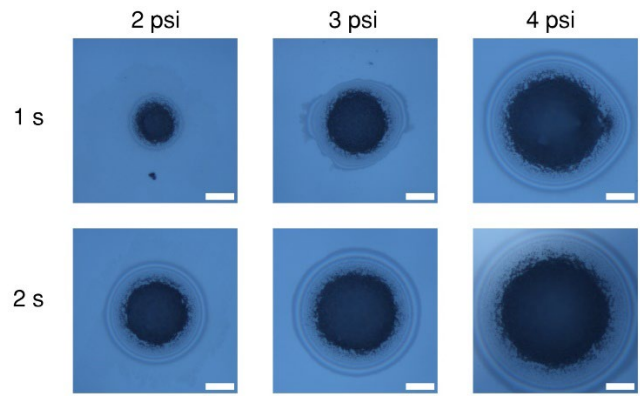

**Supplementary Figure 1. Microscope images of deposited carbon black and PDMS composite.**

The coverage area was controlled through tuning the injection pressure and time. Injection time was varied between 1 second and 2 seconds, and the pressure was varied from 2 psi, 3 psi and 4 psi. Scale bar: 50 μm.

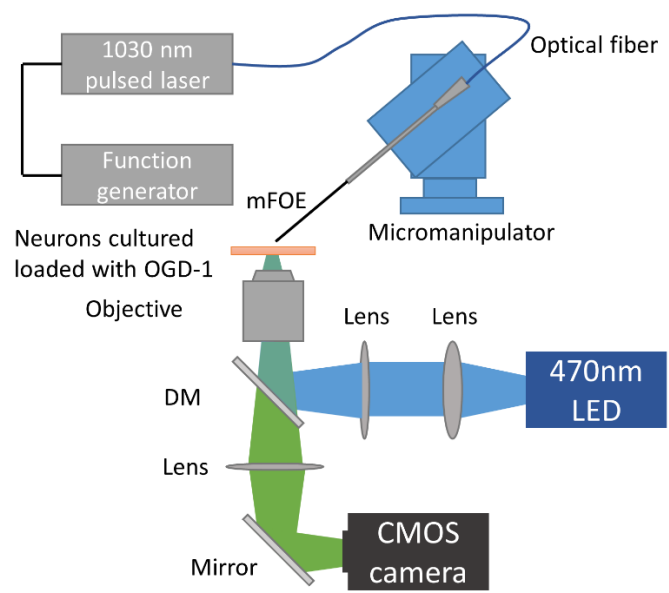

**Supplementary Figure 2. Schematic of in vitro mFOE stimulation and calcium imaging set up.**

Stimulation: 1030 nm pulsed laser is triggered by a function generator and delivered to the mFOE through an optical fiber. Calcium imaging: Oregon green is excited by 470 nm LED and the fluorescence signal is detected through a CMOS camera.

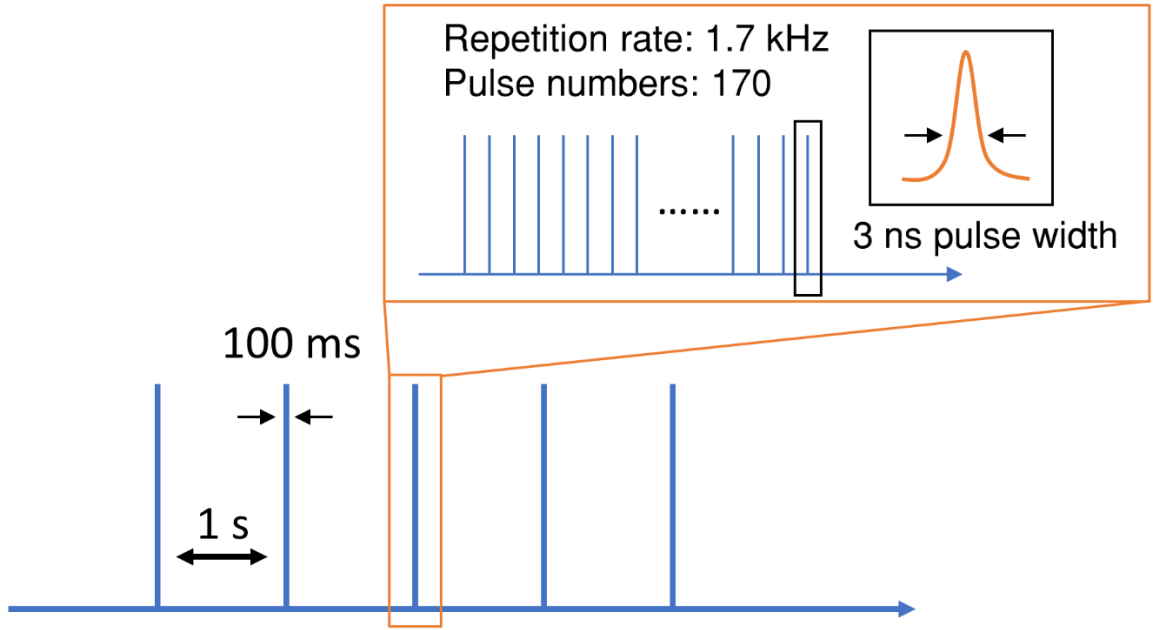

**Supplementary Figure 3. Illustration of the laser pulse train for 5 bursts with 100 ms duration at 1Hz.**

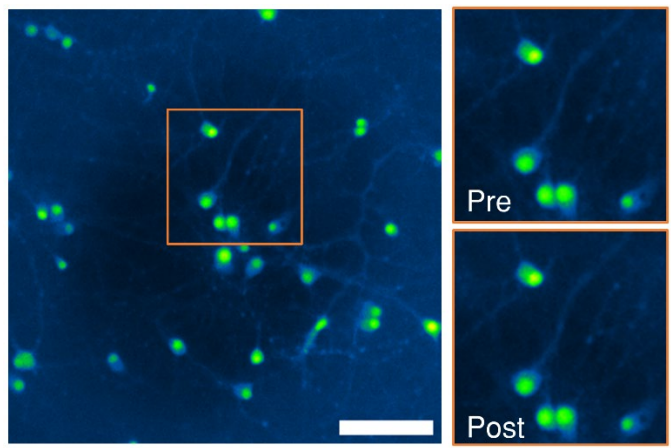

**Supplementary Fig. 4 Calcium imaging of neurons before and after mFOE stimulation.** Scale bar: 100 μm.

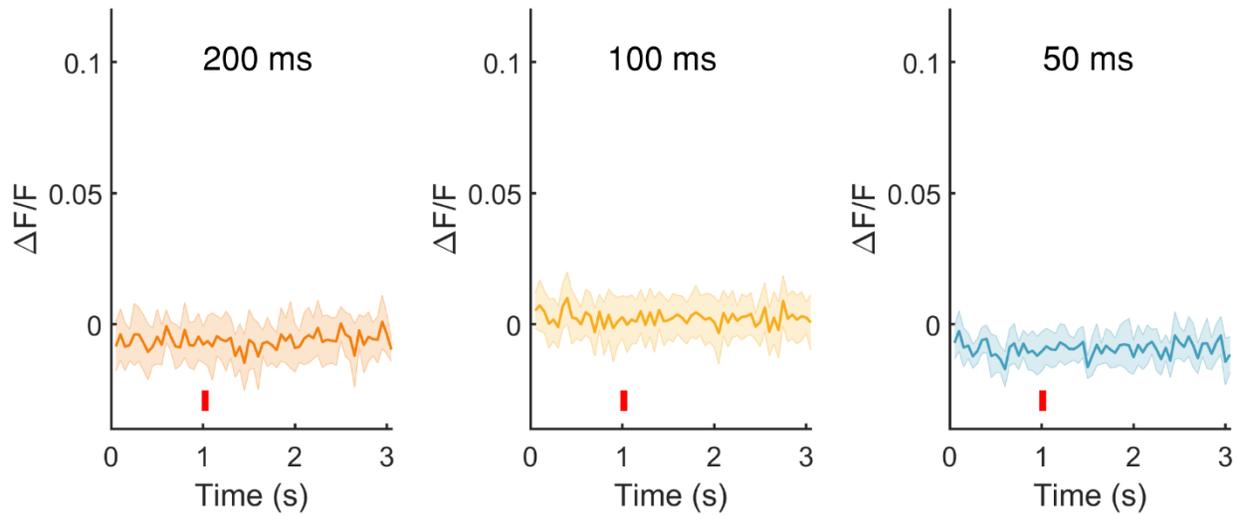

**Supplementary Fig. S5 Average calcium traces of laser only control groups.** The laser duration was same with three conditions tested in mFOE stimulation (200 ms, 100 ms and 50 ms). Laser light with pulse energy of 41.8 μJ was triggered at the time point labelled by the red bar. Shaded areas: standard deviation. (N=3)

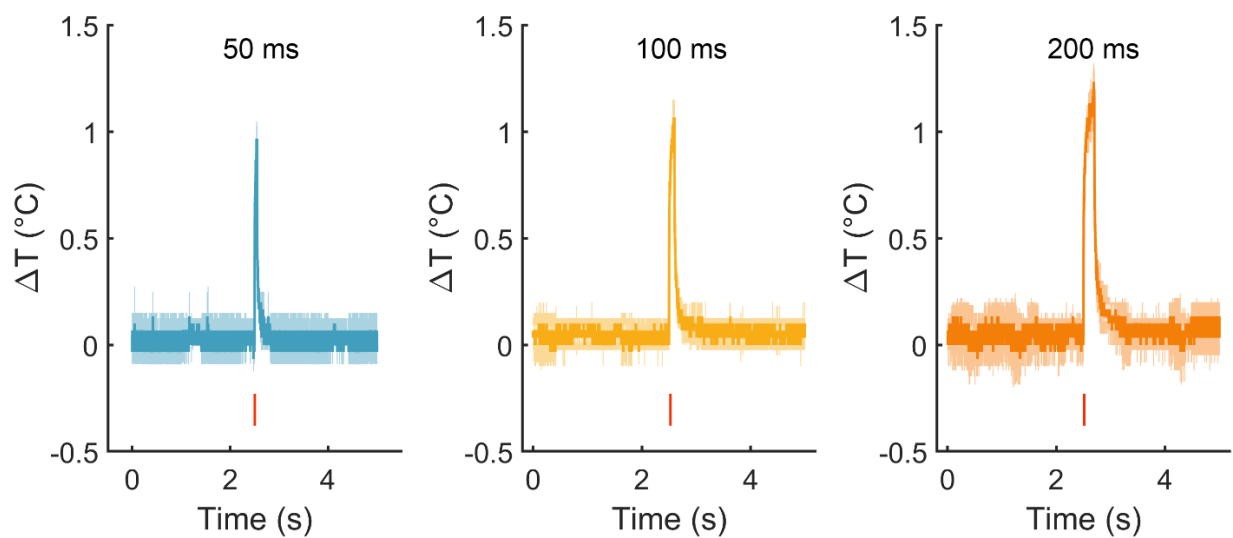

**Supplementary Fig. S6 Temperature change of the optoacoustic emitter integrated on mFOE.** The pulse energy was maintained at 41.8 µJ and the burst duration was varied from 50 ms (blue), 100 ms (yellow) to 200 ms (orange). Laser was trigger at 2.5 second as labelled by the red bar.

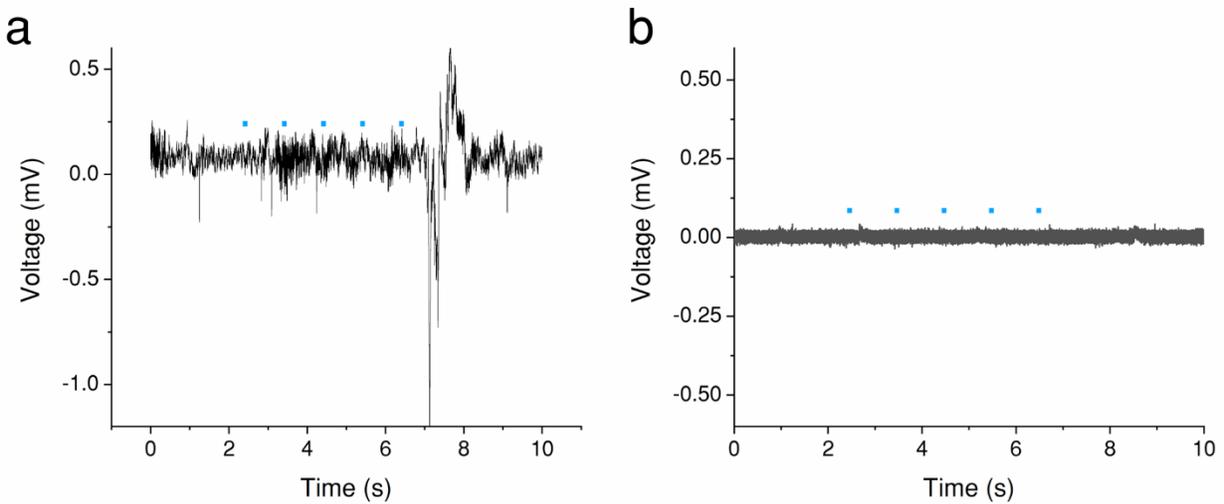

**Supplementary Fig. S7 LFP recording of sham control stimulation experiments.**

**a.** Electrophysiological recording under light only stimulations delivered through a bare multifunctional fiber without optoacoustic emitter. **b.** Simultaneous optoacoustic stimulation and electrophysiological recording of an euthanized mouse. Same laser condition was used: pulse energy of 41.8 µJ, 50 ms burst of pulses, 1 Hz, blue dots indicate the laser onset.